\newcommand{\stdpack}{
  \usepackage{amssymb}
  \usepackage{amsmath}
  \usepackage{amsfonts}
  \usepackage{amsthm}
  \usepackage{eucal}
  \usepackage[final]{graphicx}
  \usepackage{psfrag}
  \usepackage{fancyhdr}
  \renewcommand{\headrulewidth}{.5pt}\renewcommand{\footrulewidth}{.0pt}\cfoot{}
  \fancyhead[OR]{\it\theauthor---\today}
  \fancyhead[OL]{\rightmark}
  \fancyhead[ER]{\leftmark}
  \fancyhead[EL]{}
  \fancyfoot[EL,OR]{\thepage}

  \theoremstyle{plain}
  \newtheorem{theorem}{Theorem}[section]
  \newtheorem{lemma}[theorem]{Lemma}
  \newtheorem{corollary}[theorem]{Corollary}

  \theoremstyle{definition}
  \newtheorem{definition}{Definition}[section]
  \theoremstyle{remark}

  \newcommand{\draft}{\usepackage[light,first]{draftcopy}\draftcopyName{draft}{350}}
  \newcommand{\labels}{\usepackage{/home/mt/tex/showlabels}}
  \newcommand{\maple}{\usepackage{maple2e}}
  \newcommand{\makeidx}{\usepackage{makeidx}\makeindex}
  \newcommand{\chicago}{\usepackage{chicago}\bibliographystyle{chicago}}
  \newcommand{\pdflatex}{
    \usepackage{color}
    \definecolor{urlcol}{rgb}{0,0,.6}
    \definecolor{citecol}{rgb}{0,.6,0}
    \usepackage[
    urlcolor=urlcol,
    citecolor=citecol,
    pdftex,
    colorlinks,
    backref,
    pagebackref,
    pdfpagemode=None,
    pdftitle={Authors and References},
    pdfauthor={Marc Toussaint},
    pdfstartview=FitH
    ]{hyperref}
  }
}
\newcommand{\cleardefs}{
  \renewcommand{\article}[2]{}
  \renewcommand{\book}[2]{}
  \renewcommand{\draft}{}
  \renewcommand{\labels}{}
  \renewcommand{\maple}{}
  \renewcommand{\makeidx}{}
  \renewcommand{\chicago}{}
  \renewcommand{\pdflatex}{}
  \renewcommand{\header}{}
}
\newcommand{\stdstyle}[1]{
  \stdpack
  \usepackage{./mt}
  \newcommand{\blockindent}{3ex}
  \renewcommand{\baselinestretch}{#1}
  \renewcommand{\arraystretch}{1.2}

  \usepackage{geometry}
  \geometry{a4paper,dvips,hdivide={20mm,*,20mm},vdivide={30mm,*,20mm},twosideshift=0mm}

  \columnsep 5ex
  \parindent 3ex
  \parskip 1ex
  \pagestyle{fancy}
  \renewenvironment{abstract}
  {\begin{rblock}\hrule\vspace{2ex}{\bf Abstract.~}\small}
    {\vspace{3ex}\hrule\end{rblock}\vspace{5ex}}
}

\newcommand{\article}[2]{
  \documentclass[#1pt,twoside,fleqn]{article}
  \stdstyle{#2}
  \macros 
}
\newcommand{\nips}{
  \documentclass{article}
  \usepackage{nips01e,times}
  \stdpack\macros
}
\newcommand{\ijcnn}{
  \documentclass[10pt,twocolumn]{/home/mt/tex/ijcnn}
  \stdpack\macros
  \bibliographystyle{abbrv} 
}
\newcommand{\foga}{
  \documentclass{article} 
  \stdpack\macros
  \usepackage{/home/mt/tex/foga-02}
  \usepackage{/home/mt/tex/chicago}
  \bibliographystyle{/home/mt/tex/foga-chicago}
}
\newcommand{\book}[2]{
  \documentclass[#1pt,twoside,fleqn]{book}
  \newenvironment{abstract}{\begin{rblock}{\bf Abstract.~}\small}{\end{rblock}}
  \stdstyle{#2}
  \macros
}

\newcommand{\foils}[1]{
  \documentclass[12pt,fleqn]{article}
  \stdstyle{#1}
  \voffset -1cm  \addtolength{\textheight}{2cm}
  \renewcommand{\footskip}{2cm}
  \macros
  
\begin{document}
  \large
}
\newcommand{\landslides}[1]{
  \documentclass{article}
  \stdpack
  \renewcommand{\baselinestretch}{#1}
  \renewcommand{\arraystretch}{1}

  \usepackage[landscape,a4paper,pdftex]{geometry}

  \newcommand{\pageheight}{209.9mm}
  \setlength{\voffset}{-1in}\addtolength{\voffset}{1mm}% for the printer
  \setlength{\topmargin}{2mm}
  \setlength{\headheight}{30mm}
  \setlength{\headsep}{15mm}
  \setlength{\footskip}{15mm}
  \setlength{\textheight}{\pageheight}\addtolength{\textheight}{-70mm}

  \newcommand{\pagewidth}{297.0mm}
  \setlength{\hoffset}{-1in}\addtolength{\hoffset}{1mm}
  \setlength{\oddsidemargin}{20mm}
  \setlength{\textwidth}{\pagewidth}\addtolength{\textwidth}{-40mm}

  \setlength{\columnsep}{30mm}
  \parindent 0ex
  \parskip 0ex
  \setlength{\itemsep}{8ex}
  \pagestyle{fancy}
  \renewcommand{\headrulewidth}{3pt}
  \renewcommand{\footrulewidth}{1pt}

  \renewcommand{\labelenumi}{\textbf{\arabic{enumi}.}~~}

  \newcommand{\theslide}{~}
  \newcommand{\theauthor}{Marc Toussaint}
  \rhead{\Large \thetitle\\ \it \theauthor}
  \lhead{
    \inilogo
    {\Huge\sc \quad\theslide}\\
  }
  \rfoot{\thepage}

  \macros
}
\newcommand{\landfolien}[1]{
  \documentclass[fleqn]{article}
  \usepackage{german}
  \stdpack
  \renewcommand{\baselinestretch}{#1}
  \renewcommand{\arraystretch}{1.5}

  \usepackage[landscape,a4paper,pdftex]{geometry}

  \setlength{\hoffset}{-5cm}%{-6cm}
  \setlength{\voffset}{-1.5cm}
  \setlength{\textwidth}{27cm}%{29cm}
  \setlength{\textheight}{19cm}
  \parindent 0ex
  \parskip 0ex %8ex
  \pagestyle{plain}
  \begin{document}
  \huge
}

\newcommand{\firstslide}[1]{\renewcommand{\theslide}{#1}}
\newcommand{\newslide}[1]{\onecolumn\renewcommand{\theslide}{#1}}
\newcommand{\newslidetwo}[1]{\twocolumn\renewcommand{\theslide}{#1}}

%------------------------------------------------------------------------------
% title page

\author{Marc Toussaint}

\newcommand{\inilogo}{
}

\newcommand{\addressCologne}{
  Institute for Theoretical Physics\\
  University of Cologne\\
  50923 K\"oln---Germany\\
  {\tt mt@thp.uni-koeln.de}\\
  {\tt www.thp.uni-koeln.de/\~{}mt/}
}

\newcommand{\homepage}{{\tt www.neuroinformatik.ruhr-uni-bochum.de/PEOPLE/mt/}}
\newcommand{\email}{{\rm mt@neuroinformatik.ruhr-uni-bochum.de}}

\newcommand{\address}{
  Institut f\"ur Neuroinformatik,
  Ruhr-Universit\"at Bochum, ND 04,
  44780 Bochum---Germany\\
  \email
  %\homepage
}

\newcommand{\published}{}

\newcommand{\mytitle}{
  \thispagestyle{empty}
  \hrule height2pt
  \begin{list}{}{\leftmargin2ex \rightmargin2ex \topsep2ex }\item[]
    {\huge\bf \thetitle}
  \end{list}
  \begin{list}{}{\leftmargin7ex \rightmargin7ex \topsep0ex }\item[]
    Marc Toussaint \quad\today

    {\small\it \address}
  \end{list}
  \vspace{2ex}
  \hrule height1pt
  \vspace{5ex}
  \renewcommand{\mytitle}{\chapter{\thetitle}}
}
\renewcommand{\mytitle}{
  \thispagestyle{empty}
  \mbox{~}
  \begin{list}{}{\leftmargin4ex \rightmargin4ex \topsep17ex }\item[]
    \center
    {\LARGE \thetitle}

    \vspace{7ex}
    {\large \theauthor}

    \vspace{1ex}
    {\footnotesize\sl \address}

    {\footnotesize\today}

    \vspace{1ex}
    {\small \published}
  \end{list}
  \vspace{8ex}
  \renewcommand{\mytitle}{\chapter{\thetitle}}
}

%------------------------------------------------------------------------------
% environments / commands

\newcommand{\Section}[2]{
  \section*{\refstepcounter{section}\thesection\quad #1}
  \addcontentsline{toc}{section}
  {\protect\numberline{\thesection}#1\newline{\it #2}}
  \begin{rblock}\it #2\end{rblock}\medskip
}
\newcommand{\Subsection}[3]{
  \subsection*{\refstepcounter{subsection}\thesubsection\quad #2}\label{#1}
  \addcontentsline{toc}{subsection}
  {\protect\numberline{\thesubsection}#2
  \newline{\protect\begin{minipage}[t]{3ex}~\protect\end{minipage}
           \protect\begin{minipage}[t]{13cm}\it #3\protect\end{minipage}}
  }
  \begin{rblock}\it #3\end{rblock}\medskip
}

\newcommand{\content}[1]{
%  \begin{rblock}\it #1\end{rblock}\medskip
%  \addtocontents{toc}{\protect\begin{list}{}{\leftmargin9ex
%        \rightmargin9ex \topsep-2ex \parsep.5ex}}
%  \addtocontents{toc}{\protect\item[] \protect\small\protect\it #1}
%  \addtocontents{toc}{\protect\end{list}\protect\medskip}
}

\newcommand{\sepline}{
  \begin{center} \begin{picture}(200,0)
    \line(1,0){200}
  \end{picture}\end{center}
}

\newcommand{\horline}{\hrule}

\newcommand{\sepstar}{
  \begin{center} {\vspace{0.5ex}*} \end{center}\vspace{-1.5ex}\noindent
}

\newcommand{\partsection}[1]{
  \vspace{5ex}
  \centerline{\sc\LARGE #1}
  \addtocontents{toc}{\contentsline{section}{{\sc #1}}{}}
}

\newcommand{\intro}[1]{\textbf{#1}\index{#1}}

\newcounter{parac}
\newcommand{\para}{\noindent\refstepcounter{parac}{\bf [{\roman{parac}}]}~~}
\newcommand{\Pref}[1]{[\emph{\ref{#1}}\,]}

\newenvironment{block}[1][]{{\noindent\bf #1}
\begin{list}{}{\leftmargin\blockindent \topsep-\parskip}
\item[]
}{
\end{list}
}

\newenvironment{rblock}{
\begin{list}{}{\leftmargin\blockindent \rightmargin\blockindent \topsep-\parskip}
\item[]
}{
\end{list}
}

\newenvironment{algorithm}{
\begin{list}{\raisebox{.3ex}{\footnotesize\bf(\arabic{enumi})}}
{\usecounter{enumi} \leftmargin7ex \rightmargin7ex \labelsep1ex
  \labelwidth5ex \topsep0ex \parsep.5ex \itemsep0pt}
}{
\end{list}
}

\newenvironment{keywords}{\paragraph{Keywords}\begin{rblock}\small}{\end{rblock}}

\newenvironment{colpage}{
\addtolength{\columnwidth}{-3ex}
\begin{minipage}{\columnwidth}
\vspace{.5ex}
}{
\vspace{.5ex}
\end{minipage}
}

\newenvironment{enum}{
\begin{list}{}{\leftmargin3ex \topsep0ex \itemsep0ex}
\item[\labelenumi]
}{
\end{list}
}

\newenvironment{cramp}{
\begin{quote} \begin{picture}(0,0)
        \put(-5,0){\line(1,0){20}}
        \put(-5,0){\line(0,-1){20}}
\end{picture}
}{
\begin{picture}(0,0)
        \put(-5,5){\line(1,0){20}}
        \put(-5,5){\line(0,1){20}}
\end{picture} \end{quote}
}

%------------------------------------------------------------------------------
% macros

\newcommand{\macros}{
  \newcommand{\0}{{\hat 0}}
  \newcommand{\1}{{\hat 1}}
  \newcommand{\2}{{\hat 2}}
  \newcommand{\3}{{\hat 3}}
  \newcommand{\5}{{\hat 5}}

  \renewcommand{\a}{\ensuremath\alpha}
  \renewcommand{\b}{\beta}
  \renewcommand{\c}{\gamma}
  \renewcommand{\d}{\delta}
    \newcommand{\D}{\Delta}
    \newcommand{\e}{\epsilon}
    \newcommand{\g}{\gamma}
    \newcommand{\G}{\Gamma}
  \renewcommand{\l}{\lambda}
  \renewcommand{\L}{\Lambda}
    \newcommand{\m}{\mu}
    \newcommand{\n}{\nu}
    \newcommand{\N}{\nabla}
  \renewcommand{\k}{\kappa}
  \renewcommand{\o}{\omega}
  \renewcommand{\O}{\Omega}
    \newcommand{\p}{\phi}
    \newcommand{\ph}{\varphi}
  \renewcommand{\P}{\Phi}
  \renewcommand{\r}{\varrho}
    \newcommand{\s}{\sigma}
    \newcommand{\Si}{\Sigma}
  \renewcommand{\t}{\theta}
    \newcommand{\T}{\Theta}
  \renewcommand{\v}{\vartheta}
    \newcommand{\x}{\xi}
    \newcommand{\X}{\Xi}
    \newcommand{\Y}{\Upsilon}

  \renewcommand{\AA}{{\cal A}}
    \newcommand{\GG}{{\cal G}}
  \renewcommand{\SS}{{\cal S}}
    \newcommand{\TT}{{\cal T}}
    \newcommand{\EE}{{\cal E}}
    \newcommand{\FF}{{\cal F}}
    \newcommand{\HH}{{\cal H}}
    \newcommand{\II}{{\cal I}}
    \newcommand{\KK}{{\cal K}}
    \newcommand{\LL}{{\cal L}}
    \newcommand{\MM}{{\cal M}}
    \newcommand{\NN}{{\cal N}}
    \newcommand{\CC}{{\cal C}}
    \newcommand{\OO}{{\cal O}}
    \newcommand{\PP}{{\cal P}}
    \newcommand{\QQ}{{\cal Q}}
    \newcommand{\RR}{{\cal R}}
    \newcommand{\UU}{{\cal U}}
    \newcommand{\YY}{{\cal Y}}
    \newcommand{\SOSO}{{\cal SO}}
    \newcommand{\GLGL}{{\cal GL}}

  \newcommand{\NNN}{{\mathbb{N}}}
  \newcommand{\ZZZ}{{\mathbb{Z}}}
  \newcommand{\RRR}{{\mathbb{R}}}
  \newcommand{\CCC}{{\mathbb{C}}}
  \newcommand{\one}{{{\bf 1}}}

  \renewcommand{\[}{\Big[}
  \renewcommand{\]}{\Big]}
  \renewcommand{\(}{\Big(}
  \renewcommand{\)}{\Big)}
  \renewcommand{\|}{\big|}
  \newcommand{\<}{{\ensuremath\langle}}
  \renewcommand{\>}{{\ensuremath\rangle}}

  \newcommand{\Prob}{{\rm Prob}}
  \newcommand{\Aut}{{\rm Aut}}
  \newcommand{\cor}{{\rm cor}}
  \newcommand{\corr}{{\rm corr}}
  \newcommand{\cov}{{\rm cov}}
  \newcommand{\sd}{{\rm sd}}
  \newcommand{\tr}{{\rm tr}}
  \newcommand{\Tr}{{\rm Tr}}
  \newcommand{\id}{{\rm id}}
  \newcommand{\Gl}{{\rm Gl}}
  \newcommand{\lag}{\mathcal{L}}
  \newcommand{\inn}{\rfloor}
  \newcommand{\lie}{\pounds}
  \newcommand{\longto}{\longrightarrow}
  \newcommand{\speer}{\parbox{0.4ex}{\raisebox{0.8ex}{$\nearrow$}}}
  \renewcommand{\dag}{ {}^\dagger }
  \newcommand{\h}{{}^\star}
  \newcommand{\w}{\wedge}
  \newcommand{\too}{\longrightarrow}
  \newcommand{\To}{\Rightarrow}
  \newcommand{\Too}{\;\Longrightarrow\;}
  \newcommand{\ow}{\stackrel{\circ}\wedge}
  \newcommand{\feed}{\nonumber \\}
  \newcommand{\comma}{\; , \quad}
  \newcommand{\period}{\; . \quad}
  \newcommand{\del}{\partial}
  \newcommand{\point}{$\bullet~~$}
  \newcommand{\doubletilde}{
  ~ \raisebox{0.3ex}{$\widetilde {}$} \raisebox{0.6ex}{$\widetilde {}$} \!\!
  }
  \newcommand{\topcirc}{\parbox{0ex}{~\raisebox{2.5ex}{${}^\circ$}}}
  \newcommand{\sym}{\topcirc}

  \newcommand{\half}{\frac{1}{2}}
  \newcommand{\third}{\frac{1}{3}}
  \newcommand{\fourth}{\frac{1}{4}}

  \newcommand{\ubar}{\underline}

  \renewcommand{\_}{\underset}
  \renewcommand{\^}{\overset}
  \renewcommand{\*}{{\rm\raisebox{-.6ex}{\text{*}}{}}}
}

\newcommand{\RND}{{\SS}}
\newcommand{\IF}{\text{if }}
\newcommand{\AND}{\textsc{and }}
\newcommand{\OR}{\textsc{or }}
\newcommand{\XOR}{\textsc{xor }}
\newcommand{\NOT}{\textsc{not }}
\newcommand{\argmax}[1]{\text{arg}\!\max_{#1}}
\newcommand{\argmin}[1]{\text{arg}\!\min_{#1}}
\newcommand{\ee}[1]{\ensuremath{\,\cdot10^{#1}}}
\newcommand{\sub}[1]{\ensuremath{_{\text{#1}}}}
\newcommand{\up}[1]{\ensuremath{^{\text{#1}}}}
\newcommand{\kld}[2]{D\big(#1\,\big|\!\big|\,#2\big)}
\newcommand{\sprod}[2]{\big<#1\,,\,#2\big>}
\newcommand{\End}{\text{End}}
\newcommand{\txt}[1]{\quad\text{#1}\quad}

%------------------------------------------------------------------------------
% stuff

\newcommand{\anchor}[3]{\begin{picture}(0,0)\put(#1,#2){#3}\end{picture}}

\newcommand{\pathmt}{./}
\newcommand{\basepath}{./}
\newcommand{\setpath}[1]{\renewcommand{\pathmt}{#1}\renewcommand{\basepath}{#1}}
\newcommand{\pathinput}[2]{
  \renewcommand{\pathmt}{\basepath #1}
  \input{\pathmt #2} \renewcommand{\pathmt}{\basepath}}

\newcommand{\hide}[1]{{\tt[hide:~}{\small #1}{\tt]}\message{HIDE--Warning}}
\newcommand{\color}[2][1]{}
\newcommand{\todo}[1]{{\tt[TODO: #1]}\message{TODO--Warning: #1}}
\newcommand{\header}{\begin{document}\mytitle\cleardefs}
\newcommand{\contents}{{\small\tableofcontents}}
\newcommand{\footer}{\small\bibliography{/home/mt/tex/bibs}\end{document}}
\article{10}{1}
\chicago
%\labels

\renewcommand{\Subsection}[3]{\section{#2}\label{#1}}
\newcommand{\dissopt}[2]{#2}
\newcommand{\tO}{\tilde\O}
\renewcommand{\todo}[1]{}
\renewcommand{\hide}[1]{}

\title{The structure of evolutionary exploration: On crossover,
  buildings blocks and Estimation-Of-Distribution Algorithms\\[2ex]
  DRAFT}

\author{Marc Toussaint}

\header

\begin{abstract}
  The notion of building blocks can be related to the structure of the
  offspring probability distribution: loci of which variability is
  strongly correlated constitute a building block. We call this
  \emph{correlated exploration}. With this background we 
  analyze the structure of the offspring probability distribution, or
  \emph{exploration distribution}, for a GA with mutation only, a
  crossover GA, and an Estimation-Of-Distribution Algorithm (EDA). The
  results allow a precise characterization of the structure of the
  crossover exploration distribution. Essentially, the crossover
  operator destroys mutual information between loci by transforming it
  into entropy; it does the inverse of correlated exploration. In
  contrast, the objective of EDAs is to model the mutual information
  between loci in the fitness distribution and thereby they induce
  correlated exploration.
\end{abstract}

\Subsection{crossintro}
{Introduction}{}

\dissopt{ In the previous sections we neglected crossover although the
  discussion of crossover has a very strong tradition in the field of
  genetic algorithms. So we now append a discussion of crossover
  w.r.t. correlated exploration by comparing it to mutational
  exploration and Estimation-Of-Distribution Algorithms (EDAs). This
  section will primarily contribute to the field of theoretical
  evolutionary computation.}{}

In the realm of evolutionary computation the notion of building blocks
of evolution has been developed in Holland's original works
\cite{holland:75,holland:00} to describe the effect of crossover. In
that respect, building blocks are composed of genes with more or less
linkage between them. This is one to one with the notion of schemata
and eventually lead to the schema theories which describe the
evolution of these building blocks.

In the biology literature though, the notion of building blocks has
quite a different connotation. As a paradigm I choose the empirical
findings of \citeN{halder-callaerts-gehring:95}: The experimenters
forced the mutation of a single gene, called ``eyeless gene'', in early
ontogenesis of a Drosophila Melanogaster fly. This rather subtle
genotypic variation results in a severe phenotypic variation: an
additional whole, functionally complete eye module grows at some place
it was not supposed to. Here, the notion of a building block refers to
the eye as a functional module which can be grown phenotypically by
triggering a single gene. In other words, a single (and thus
non-correlated) mutation of a gene leads to a highly complex, in terms
of physiological cell variables highly correlated phenotypic
variation. Such properties of the genotype-phenotype mapping are
considered as the basis of complex adaptation
\cite{wagner-altenberg:96}. \dissopt{}{A theory on the evolution of complex
phenotypic variability exists \cite{toussaint:03}, and in this paper
we show that the induced notion of building blocks is completely
different to the one induced by crossover.}

Besides the discussion of crossover in GAs and that of functional
modularity in natural evolution, there is a third field of research
that relates to the discussion of building blocks:
Estimation-of-Distribution Algorithms (EDAs,
\citeNP{pelikan-goldberg-lobo:99}). These algorithms are a direct
implementation of the idea of correlated exploration in the framework
of heuristic search algorithms. They explicitly encode the search
distribution (i.e., offspring probability distribution) by means of a
product of marginals (PBIL, \citeNP{baluja:94}), factorized
distributions (FDA, \citeNP{muehlenbein-mahnig-rodriguez:99}),
dependency trees (\citeNP{baluja-davies:97}), or, most generally, a
Bayesian network (BOA, \citeNP{pelikan-goldberg-cantupaz:00}). To my
point of view, the key of these algorithms is that they are capable to
induce the same notion of building blocks as we introduced it in the
context of natural evolution. For instance, consider a dependency tree
where the leaves encode the phenotypic variables. Offsprings are
generated by \emph{sampling} this probabilistic model, i.e., by first
sampling the root variable of the tree, then, according to the
dependencies encoded on the links, sampling the root's successor
nodes, etc. Now, if we assume that the dependencies are very strong,
say, deterministic, it follows that a single variation at the root
leads to a completely correlated variation of all leaves. Hence, we
may define a set of leaves which, due to their dependencies, always
vary in high correlation as a functional phenotypic module in the same
sense as for the eyeless paradigm.

Several discussions in the EC community though contradict this point
of view: Some argue that the essence of EDAs is that they can model
the evolution of crossover building blocks (schemata) by explicitly
encoding the linkage correlations that are implicit in the offspring
distribution of crossover GAs \cite[Introduction]{shapiro:03}. In that
sense, EDAs are ``only'' faster versions of crossover GAs; faster
because EDAs actively analyze correlations in the selection
distribution whereas crossover masks would have to self-adapt (see
section \ref{crossconclu}). In this \dissopt{section}{paper} we want
to point out that, certainly, crossover induces a correlation in the
search distribution that can be modeled by graphical models, but the
concept of graphical models is far more general than that of linkage
correlations.  Hence, EDAs and non-trivial gene interaction models
(non-trivial genotype-phenotype mappings\dissopt{}{,
  \citeNP{toussaint:03}}) can introduce correlational structures in
the search distribution that go \emph{qualitatively} beyond simple
crossover GAs.

Most important of all: EDAs and gene interaction models can account
for \emph{correlated innovation}. Here, innovation means that some
phenotypic variable changes its value and some other phenotypic
variables change their values in high dependence of this change, such
that the constellation of this set of variables is really new, has not
been present in the parent population. In contrast, crossover can only
preserve certain (by the crossover mask determined) linkage
correlations that have been present in the parent population and never
explores new correlated constellations in the sense of correlated
innovation.

\begin{figure}\center
\input{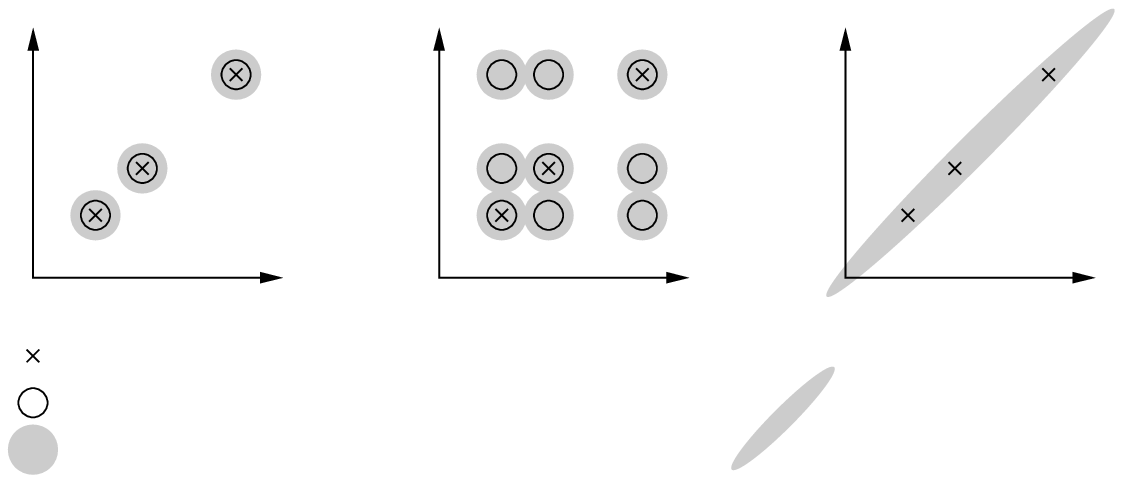}\\
\caption{\label{corr}
  Illustration of the type of correlations in GAs with and without
  crossover in comparison to correlated exploration in EDAs. The gray
  shades indicate the exploration distributions, say, regions of
  probability greater than some constant. The degree to which the gray
  shading is aligned with the bisecting line indicates correlatedness.
  The crossover GA in the middle destroys correlations whereas EDAs
  may induce high correlations.}
\end{figure}

The main goal of this \dissopt{section}{paper} is to prove and
formalize the claims that have been made above. After we define
crossover in the next section, section \ref{crossMut} and
\ref{crossinfo} will present some theorems on the `structure' of the
search distribution after mutation and crossover. With structure we
mean the correlational structure that we measure by means of mutual
information. Many arguments are based on the increase and decrease of
mutual information in relation to increase or decrease of entropy in
the search distribution. Section \ref{correxplo} finally defines the
notion of correlated exploration and thereby pinpoints the difference
between linkage correlations and correlations in EDAs or gene
interaction models. Figure \ref{corr} already explains the key idea.

\Subsection{crossform}
{Formalism}{}

\paragraph{The Simple GA.}

We represent a population as a distribution $p$ over genotype space
$\O$. In this paper we assume that a genotype is composed of a fixed
number of genes, $\O = \O^1 \times \cdots \times \O^N$, where the
space $\O^i$ of alleles of the $i$th gene is arbitrary. We represent
also finite populations as a distribution $p \in \L^\O$, namely, if
the population is given as a multiset $A=\{x_1,..,x_\m\}$ we
(isomorphically) represent it as the \emph{finite distribution} given
by $p=\frac{1}{\m} \sum_{i=1}^\m \d_{x_i}$ where $\d_x$ is the delta
distribution at $x$, i.e., $p(x) = \frac{|A \cap \{x\}|}{|A|} =
\frac{\text{multiplicity of $x$ in $A$}}{|A|}$. Crossover and mutation
are represented as operators $\L^\O \to \L^\O$ that map a parental
(finite or infinite) population to an offspring distribution. Given
some operator $\UU:\, \L^\O \to \L^\O$ we will use the notation
$\D_\UU B = B(\UU p) - B(p)$ to denote the difference of a quantity
$B:\, \L^\O \to \RRR$ under transition, e.g., the quantity may be the
entropy $H(p)$ of a distribution.

In that framework we may write the evolution equation of a crossover
GA as
\begin{align*}
p^{(t+1)} = \SS^\m\, \FF^{(t)}\, \SS^\l\, \MM\, \CC\, p^{(t)} \;,
\end{align*}
with crossover $\CC$, mutation $\MM$, offspring sampling $\SS^\l$,
fitness $\FF$, and parent sampling $\SS^\m$. A sampling operator
$\SS^n:\, \L^\O \to \L^\O$ draws $n$ independent samples from a
distribution and maps this multiset of samples to the respective
finite distribution; note that $\lim_{n\to\infty} \SS^n = \id$.
Fitness $\FF^{(t)}:\, \L^\O \to \L^\O$ rescales a distribution
proportional to some functional $f^{(t)}, (\FF^{(t)} p)(x) =
\frac{f^{(t)}(x)\, p(x)}{\sum_{x'} f^{(t)}(x')\, p(x')}$. We define
mutation and crossover more precisely as follows:

\begin{definition}[Mutation]\label{defMut}
  We define mutation as an operator $\MM:\, \L^\O \to \L^\O$ defined
  by the conditional probability $\MM(y|x)$ of mutating from $x\in\O$
  to $y\in\O$:
\begin{align*}
\MM p = \sum_x \MM(\cdot|x)\, p(x) \;.
\end{align*}
A \emph{typical} mutation operator fulfills the constraints of symmetry
and component-wise independence:
\begin{align*}
&\text{a)}\qquad \MM(y|x) = \MM(x|y) \\
&\text{b)}\qquad \O = \O^1 \times \cdots \times \O^N
 ~\To~ \MM(x|y) = \prod_{i=1}^N \MM^i(x^i|y^i)
\end{align*}
In the following we will refer to the the \emph{simple} mutation
operator for which all component-wise mutation operators are such that
the probability of mutating from $x$ to $y$ is constant for
$x\not=y$:
\begin{align*}
  \forall i: \MM^i = \MM^* \comma
  \forall x\not= y \in \O^*:\; \MM^*(x|y) = \frac{\a}{n} \comma
  \forall x\in \O^*:\; \MM^*(x|x) = 1-\frac{\a\, (n-1)}{n} \;,
\end{align*}
where $n=|\O^*|$ and $0\le\a\le1$ denotes the mutation rate parameter.
\end{definition}

\begin{definition}[Crossover]
  We define crossover as an operator $\L^\O \to \L^\O$ parameterized
  by a mask distribution $c \in \L^{\{0,1\}^N}$, where $N$ is the
  number of loci (or genes) of a genome in $\O$:
\begin{align*}
\CC:\, \L^\O \to \L^\O:\,
& p \mapsto \CC p = \sum_{x_0,x_1 \in \O} \CC(\cdot | x_0,x_1)\, p(x_0)\, p(x_1) \;,\\
& \CC(x | x_0,x_1) = \sum_{m \in \{0,1\}^N} c(m) \, [x=x_0 \otimes_m x_1] \;,
  \end{align*}
  where the $i$th allele of the $m$-crossover-product $x_0 \otimes_m
  x_1$ is the $i$th allele of the parent $x_{m_i}$, i.e., $(x_0
  \otimes_m x_1)^i = (x_{m_i})^i$. We only consider \emph{symmetric
    crossover}, where $c(m)=c(\bar m)$.
\end{definition}

\noindent In the case of bit strings, $\O = \{0,1\}^N$, it holds $x_0
\otimes_m x_1 = x_0 \otimes m \oplus \bar m \otimes x_1$, where
$\oplus$ denotes the \XOR and $\otimes$ the \AND. It follows
that \cite[Theorem 4.4]{vose:99}
\begin{align*}
\forall x_0,x_1 \in \O:~ \CC(\cdot | x_0,x_1) = \CC(\cdot | x_1,x_0) \comma
\CC(x | x_0,x_1) = \CC(0 | x_0 \oplus x,x_1 \oplus x) \;.
\end{align*}

\todo{discuss that $\CC$ and $\MM$ are stochastic operators instead of
  sampling ones}

\paragraph{Estimation-Of-Distribution Algorithms.}
Concerning EDAs, we write their dynamics as
\begin{align*}
y^{(t+1)} = \HH \(\FF\, \SS^\l\, \P y^{(t)} \,,\, y^{(t)} \) \;,
\end{align*}
where, instead of a parent population, some other parameters $y^{(t)}$
(e.g. a Bayesian graph or dependency tree) determine the offspring
distribution $\P y^{(t)}$, which is sampled, evaluated, and, instead
of a simple parent sampling, mapped back on new parameters $y^{(t+1)}$
by some update operator $\HH$. The operator $\HH$ is called
\emph{heuristic rule} and, in the case of Estimation-of-Distribution
Algorithms, is such that the new search distribution $\P y^{(t+1)}$
approximates the experienced fitness distribution $\FF\, \SS^\l\, \P
y^{(t)}$. The generic implementation of this idea is
\begin{align*}
y^{(t+1)} = y^* \comma
y^* = \argmin{y \in Y} \kld{\FF\, \SS^\l\, \P y^{(t)}}{\P y} \;,
\end{align*}
where $Y$ is the space of feasible parameters $y$ and
$\kld{\cdot}{\cdot}$ denotes the Kullback-Leibler distance (see
\citeN{toussaint:03b} for a discussion of generic heuristic search
and evolution). In fact, the BOA algorithm
\cite{pelikan-goldberg-cantupaz:00}, which uses Bayesian networks to
parameterize the search distribution, realizes exactly this scheme.
Other algorithms
\cite{baluja-davies:97,muehlenbein-mahnig-rodriguez:99,baluja:94}
differ in some details, e.g., they use distance measures other than
the Kullback-Leibler divergence or realize a gradual adaptation of
continuous parameters $y$ of the style ``$y^{(t+1)} = a\, y^* +
(1-\a)\, y^{(t)}$''. \dissopt{}{See \cite{toussaint:03} for a survey on the
relation between EDAs and the evolution of genetic representations
(\emph{$\s$-evolution}) in the context of non-trivial
genotype-phenotype mappings.}

\Subsection{crossMut}
{The structure of the mutation distribution}{}

This section derives a theorem that simply states that mutation
increases entropy and decreases mutual information. It is surprising
how non-trivial it is to prove this intuitively trivial statement.

\begin{lemma}[Component-wise mutation]
  Consider the component-wise simple mutation operator $\MM^*$ as
  given in definition \ref{defMut}. It follows that
\begin{itemize}
\item[a)]
\begin{align*}
\MM^* p(x) = (1-\a)\, p(x) - \a\, \frac{1}{n} \;,
\end{align*}
which is a linear mixture between $p$ and the uniform distribution
{\rm (``$\frac{1}{n}$'')} with mixture parameter $\a$.

\item[b)] For every non-uniform population $p$, the entropy of $\MM^* p$
  is greater than the entropy of $p$,
\begin{align*}
H(\MM^* p) > H(p) \;.
\end{align*}
\end{itemize}
\end{lemma}

\begin{proof} a)
\begin{align*}
\MM^* p(x)
&= \sum_y \MM^*(x|y)\, p(y)
 = \sum_y \frac{\a}{n}\, p(y) +\[\(1-\frac{\a\, (n-1)}{n}\)-\frac{\a}{n}\]\, p(x)
 = \frac{\a}{n} + (1-\a)\, p(x) \;.
\end{align*}

b) We generally show that the entropy increases if you mix a
  distribution with the uniform distribution. We prove this by
  considering the first two derivatives of the entropy functional
  with respect to the mixture parameter $\a$. Let
\begin{align*}
&q(x) = (1-\a)\, p(x) +  \frac{\a}{n} \;,
\end{align*}
and recall $H(q) = -\sum_x q(x) \ln q(x)$ and $(X \ln X)' = X' ((\ln
X) + 1)$. It follows
\begin{align*}
& \frac{\del}{\del \a} H(q)
= - \sum_x \[-p(x)+\frac{1}{n}\] (\ln q(x) + 1)
= \sum_x \[p(x) - \frac{1}{n}\] \ln q(x) \;,\\
& \frac{\del}{\del \a} H(q) \|_{\a=1}
= \sum_x \[p(x) - \frac{1}{n}\] \ln \frac{1}{n} = 0 \;,\\
& \frac{\del^2}{\del \a^2} H(q)
= - \sum_x \frac{(p(x) - \frac{1}{n})^2}{q(x)} < 0 \quad\text{if $p$ is non-uniform.}
\end{align*}
What we found is that (1.) the entropy is maximal for the extreme case
$\a=1$ since its derivative w.r.t.\ $\a$ at this point vanishes (of
course, this corresponds to the trivial case where $q$ becomes the
uniform distribution) and (2.) the second derivative is always
negative if $p$ is non-uniform. Hence, the plot of $H$ versus $\a$ is
comparable to an upside-down parabola with maximum at $\a=1$. It
follows that for all $\a<1$ (to the left of the maximum) the
derivative $\frac{\del}{\del \a} H(q)$ is positive. Entropy
continuously increases with $\a$. And hence, for every $0\le\a\le1$
and every non-uniform population $p$, $H(\MM^* p) > H(p)$.
\end{proof}

\begin{theorem}\label{lemMut}
  Consider the simple mutation operator $\MM(x|y) = \prod_i
  \MM^*(x^i|y^i)$ as given in definition \ref{defMut}. If $p\in\L^\O$
  is non-uniform it follows that entropy increases, $H(\MM p) > H(p)$,
  and mutual information decreases, $I(\MM p) < I(p)$.
\end{theorem}
\begin{proof}
  We first prove that the cross entropy decreases. Assuming only two
  genes, the compond mutation distributions reads
\begin{align*}
\MM p(x,y)
&= (1-\a)^2\, p(x,y) + (1-\a)\a\, p(x)\, \frac{1}{n}
    + (1-\a)\,\a\, \frac{1}{n}\, p(y) + \a^2\, \frac{1}{n}\, \frac{1}{n} \\
&= (1-\a)\, \[(1-\a)\, p(x,y) + \a\, \frac{1}{n}\, p(x)\]
     + \a\, \frac{1}{n}\, \[(1-\a)\, p(y) + \a\, \frac{1}{n}\] \\
&= (1-\a)\, q(x,y) + \a\, \frac{1}{n}\, q(y) \;,\\
&\text{ where}\quad
q(x,y) = (1-\a)\, p(x,y) + \a\, p(x)\, \frac{1}{n} \comma
q(x)=p(x) \comma q(y)=(1-\a)\, p(y) + \frac{\a}{n}
\end{align*}
We call $q$ a one-component $\a$-mixture since only in one component
the uniform distribution was mixed to $p$. This shows that the compound
distribution $\MM p$ for two genes is a one-component $\a$-mixture of
a distribution $q$, which is itself a one-component $\a$-mixture. For
compound distributions with more than two genes this will be
recursively the case and generally the mutation operator can be
expresses as concatenation of one-component $\a$-mixtures.  Hence, it
suffices when we prove that the mutual information decreases for one
such step of one-component $\a$-mixing.
  
We use the same technique of calculating derivatives with respect to
the mixture parameter to proof decreasing cross entropy. To simplify
the notation we use the abbreviations:
\begin{align*}
&A=q(x,y) \;,
&&A\|_{\a=1} = \frac{\a\, p(x)}{n} \;,
&&A' = \frac{\del}{\del \a} A = -p(x,y) + \frac{p(x)}{n} \;,
&&A'' = 0 \;, \\
&B=q(x)\, q(y) = p(x)\[(1-\a)\, p(y) + \frac{\a}{n} \]\;,
&&B\|_{\a=1} = A\|_{\a=1} \;,
&&B' = p(x)\, (-p(y) + \frac{1}{n}) \;,
&&B''=0 \;.
\end{align*}
With these abbreviations (keeping the dependencies on $x$, $y$, and $\a$
in mind) we can write:
\begin{align*}
I(q)
&= \sum_{x,y} A \ln \frac{A}{B} \\
\frac{\del}{\del\a} I(q)
&= \sum_{x,y} \[ A' \ln \frac{A}{B} + A' - \frac{A\, B'}{B} \] \\
\frac{\del}{\del\a} I(q) \|_{\a=0}
&= \sum_{x,y} A'\|_{\a=1} \ln \frac{A\|_{\a=1}}{A\|_{\a=1}}
 + \sum_{x,y} \[-p(x,y) + \frac{p(x)}{n}\]
 - \sum_{x,y} \frac{A\|_{\a=1}}{A\|_{\a=1}}\, p(x)\, (-p(y) + \frac{1}{n})
 = 0 \\
\frac{\del^2}{\del\a^2} I(q)
&= \sum_{x,y} \bigg[
   A'\, \frac{B}{A} \[\frac{A'}{B} - \frac{A\, B'}{B^2} \]
   + 0 - \frac{A'B'}{B} + \frac{A\, (B')^2}{B^2} \bigg]\\
&= \sum_{x,y}\[ \frac{(A')^2}{A} - 2\, \frac{A'\, B'}{B} +\frac{A\, (B')^2}{B^2} \]
 = \sum_{x,y}\[ \frac{(B\, A' - A B')^2}{A\, B^2} \]
 > 0
\end{align*}
So, what we found is that (1.) for $\a=1$ the cross entropy is minimal
since its derivative w.r.t.\ $\a$ at this point vanishes (of course,
this corresponds to the trivial case where $q(x,y)=p(x)\,
\frac{1}{n}$) and (2.) for all other points the second derivative is
positive. The plot of $I$ versus $\a$ is comparable to an upwards
parabola with minimum at $\a=1$. It follows that for $\a<1$ (to the
left of the minimum) the derivative $\frac{\del}{\del \a} I(q)$ is
negative and thus the cross entropy continuously decreases with
increasing $\a$.

Concerning increasing entropy, it is obvious \todo{obvious} that the marginals
of the mutation distribution $\MM p$ are simply
\begin{align*}
(\MM p)^i = \MM^* p^i \;.
\end{align*}
For the component-wise mutation operators we proved that entropy
increases (for non-zero $\a$ and non-uniform $p$) and thus $\D_\MM H^i
> 0$. Consequently,
\begin{align*}
\D_\MM H = \sum_i \D_\MM H^i - \D_\MM I > 0 \;.
\end{align*}
\end{proof}

\Subsection{crossinfo}
{The structure of the crossover distribution}{}

\emph{What is the structure of the crossover search distribution $\CC
  p$, given $p \in \L^\O$ and $c \in \L^{\{0,1\}^N}$?} The first
theorem can directly be derived from our definition of the crossover
operator. It captures the most basic properties of the crossover
operator with respect to the correlations it \emph{destroys} in the
search distribution:
\begin{theorem}\label{theo1}
  Let $H(p)$, $p^i$, $H^i(p)=H(p^i)$, and $I(p)=\sum_i H^i(p) - H(p)$
  denote the entropy, the $i$th marginal distribution, the marginal
  entropies, and the mutual information of a distribution $p$
  \dissopt{}{\todo{definieren...}}. For any crossover operator $\CC$
  and any population $p$ it holds
\begin{itemize}
\item[a)] \qquad $\forall i:~ (\CC p)^i = p^i,\; \D_\CC H^i = 0$,\quad
  i.e., the marginals and hence their entropies do not change,
\item[b)] \qquad $\D_\CC I = - \D_\CC H \le 0$,\quad i.e., the
  increase of entropy is equal to the decrease of mutual information.
\end{itemize}
\end{theorem}
\begin{proof}
  Let us first calculate the marginals after crossover. Let $a$ be an
  allele of the $i$th gene.
\begin{align*}
(\CC p)^i(a)
&= \sum_{x_0,x_1} \sum_m c(m)\, [a=(x_{m_i})^i]\, p(x_0)\, p(x_1)\;, \\
&= \sum_{x_0,x_1} \sum_{m:m_i=0} c(m)\, [a=(x_0)^i]\, p(x_0)\, p(x_1)
 + \sum_{x_0,x_1} \sum_{m:m_i=1} c(m)\, [a=(x_1)^i]\, p(x_0)\, p(x_1)\;, \\
&= p^i(a) \[\sum_{m:m_i=0} c(m)\] + p^i(a) \[\sum_{m:m_i=1} c(m)\]
 = p^i(a)  \;.
\end{align*}
Since the marginals are not changed by crossover, the marginal
entropies do not change either. Statement \emph{b)} follows from the
definition of the mutual information:
\begin{align*}
\D_\CC H + \D_\CC I
&= H(\CC p) - H(p) + I(\CC p) - I(p) \\
&= H(\CC p) - H(p) + \sum_i H^i(\CC p) - H(\CC p) - \[\sum_i H^i(p) - H(p) \]\\
&= \sum_i H^i(\CC p) - \sum_i H^i(p) = 0 \;.
\end{align*}
\todo{prove $\D_\CC H\ge0$}
\end{proof}

The following theorem makes this more concrete when focusing on two
specific genes of a genome of arbitrary length. We calculate the
mutual information between these two genes in the search distribution
$\CC p$---which is a measure for the \emph{linkage} between them.
Let it be the $i$th and $j$th gene. We use $a$ and $b$ as alleles;
$p^{ij}(a,b) = \sum_{x\in\O} [x^i=a]\, [x^j=b]\, p(x)$ denotes the
probability that the $i$th gene has allele $a$ and the $j$th gene
allele $b$. Analogously, let $c^{ij}$ be the marginal of the crossover
mask distribution with respect to the two genes, i.e., $c^{ij}(01) =
\sum_{m\in\{0,1\}^N} [m^i=0]\, [m^j=1]\, c(m)$.

\begin{theorem}\label{theo2}
  For any crossover operator $\CC$ and any population $p$ it holds:
\begin{itemize}
\item[a)] The compound distribution of two genes after crossover is
  given by
\begin{align*}
(\CC p)^{ij}(a,b) &= 2\, c^{ij}(00)\, p^{ij}(a,b) + 2\, c^{ij}(01)\, p^i(a)\, p^j(b) \;,
  \end{align*}
  i.e., a linear combination of the original compound distribution $p^{ij}(a,b)$
  and the decorrelated product distribution $p^i(a)\, p^j(b)$.
  
\item[b)] The mutual information $I(\CC p)^{ij}$ in the compound
  distribution of two specific genes is
\begin{align*}
I(\CC p)^{ij}
= \sum_{a,b} \( 2 c^{ij}(00)\, p^{ij}(a,b) + 2 c^{ij}(01) \, p^i(a) p^j(b) \)
    \ln \( 2 c^{ij}(00) \frac{p^{ij}(a,b)}{p^i(a) p^j(b)} + 2 c^{ij}(01) \) \;,
\end{align*}
\item[c)] and we have
\begin{align*}
0
~\le~ 2 c^{ij}(00)\, \(I(p)^{ij} + \ln(2 c^{ij}(00))\)
~\le~ I(\CC p)^{ij} ~\le~ I(p)^{ij} \;.
  \end{align*}
  The two left $\le$ are exact for complete crossover, $c^{ij}(00)=0$,
  $c^{ij}(01)=\half$, the right $\le$ is exact for no crossover,
  $c^{ij}(00)=\half$, $c^{ij}(01)=0$.
\end{itemize}
\end{theorem}
\begin{proof}
{a)}
\begin{align*}
\CC p^{ij}(a,b)
&= \sum_{x_0,x_1} \sum_m c(m)\, [(x_{m_0})^0 = a]\, [(x_{m_1})^1 = b]\,
   p(x_0)\, p(x_1) \\
&= \sum_{x_0,x_1} \(
   c^{ij}(00)\, [(x_0)^0 = a][(x_0)^1 = b] +
   c^{ij}(01)\, [(x_0)^0 = a][(x_1)^1 = b] + \\
&\qquad\qquad  c^{ij}(10)\, [(x_1)^0 = a][(x_0)^1 = b] +
   c^{ij}(11)\, [(x_1)^0 = a][(x_1)^1 = b] \)\, p(x_0)\, p(x_1) \\
&= 2 \sum_{x_0} c^{ij}(00)\, [(x_0)^0 = a][(x_0)^1 = b]\, p(x_0)
 + 2 \sum_{x_0,x_1} c^{ij}(01)\, [(x_0)^0 = a][(x_1)^1 = b]\, p(x_0)\, p(x_1) \\
&= 2\, c^{ij}(00)\, p^{ij}(a,b) + 2\, c^{ij}(01)\, p^i(a)\, p^j(b) \;.
\end{align*}
{b\&c)}
\begin{align*}
I(\CC p)^{ij}
&= H(\CC p^i) + H(\CC p^j) - H(\CC p)
 = H(p^i) + H(p^j) - H(\CC p) \\
&\le H(p^i) + H(p^j) - H(p) = I(p)^{ij} \\
H(\CC p)
&= - \sum_{a,b}
   \( 2\, c^{ij}(00)\, p^{ij}(a,b) + 2\, c^{ij}(01)\, p^i(a)\, p^j(b) \)
   \ln\( 2\, c^{ij}(00)\, p^{ij}(a,b) + 2\, c^{ij}(01)\, p^i(a)\, p^j(b) \)\\
&= - \sum_{a,b}
   \( 2\, c^{ij}(00)\, p^{ij}(a,b) + 2\, c^{ij}(01)\, p^i(a)\, p^j(b) \)
   \[ \ln \( 2 c^{ij}(00) \frac{p^{ij}(a,b)}{p^i(a) p^j(b)} + 2 c^{ij}(01) \)
         - \ln p^i(a) - \ln p^j(b) \] \\
&= - \sum_{a,b}
   \( 2\, c^{ij}(00)\, p^{ij}(a,b) + 2\, c^{ij}(01)\, p^i(a)\, p^j(b) \)
   \[ \ln \( 2 c^{ij}(00) \frac{p^{ij}(a,b)}{p^i(a) p^j(b)} + 2 c^{ij}(01) \)
          \] + H(p^i) + H(p^j) \\
I(\CC p)^{ij}
&= \sum_{a,b}
   \( 2\, c^{ij}(00)\, p^{ij}(a,b) + 2\, c^{ij}(01)\, p^i(a)\, p^j(b) \)
   \ln \( 2 c^{ij}(00) \frac{p^{ij}(a,b)}{p^i(a) p^j(b)} + 2 c^{ij}(01) \) \\
&\ge  \sum_{a,b}
   \( 2\, c^{ij}(00)\, p^{ij}(a,b) \)
   \ln \( 2 c^{ij}(00) \frac{p^{ij}(a,b)}{p^i(a) p^j(b)} \)
 = 2 c^{ij}(00)\, \(I(p)^{ij} + \ln(2 c^{ij}(00))\)
\end{align*}
\end{proof}

Let us summarize what we actually found in the above theorems:
\begin{itemize}
\item The marginal distributions do not change at all. There is no
  exploration w.r.t.\ the alleles of single genes.
\item The more entropy crossover introduces in a population, the more
  the mutual dependencies between genes are destroyed. Actually,
  crossover destroys mutual information in the parent population by
  \emph{transforming} it into entropy in the crossed population. In
  particular, if there is no mutual information in the parent
  population, crossover will not generate any more entropy. That's
  linkage equilibrium.
\item The last theorem shows how the crossover mask distribution $c$
  determines \emph{which} correlations in the parent population are
  destroyed and transformed intro entropy.
\end{itemize}
The purpose of these theorems is to propose a probably nonstandard
point of view on what crossover actually does: It seems misleading to
say that crossover introduces the notion of building blocks. Actually,
a non-crossover GA comprises the strongest and most natural building
blocks; individuals as such are the building blocks that carry the
mutual information between their genes.  Crossover is a means to break
these maximal building blocks apart into smaller pieces by converting
mutual dependencies into entropy. As a result it induces smaller, more
fine-grained building blocks with, in total, less mutual information
in the crossed population. It is thus questionable to state that the
correlational structure in the crossed population is more complex with
crossover---actually it is simpler since it carries less information.
In the limit of linkage equilibrium (or uniform $c$), all correlations
have been destroyed and the crossed population becomes a product
distribution.

\begin{figure}
\center\input{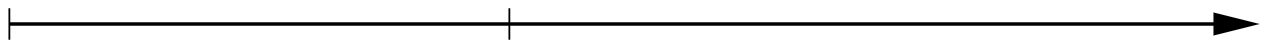}\\
\caption{\label{range}
  The degree of mutual information carried over to the exploration
  distribution. $I(p)$ is the mutual information contained in the
  parent population. Gene pool GAs, crossover, and mutation destroy
  correlations. EDAs and evolution in the case complex
  genotype-phenotype mappings (see Toussaint 2003) may induce
  correlated exploration.}
\end{figure}

\Subsection{correxplo}
{Correlated exploration}{}

What crossover and EDAs share is that both introduce a non-trivial
correlational structure in the search distribution. The crucial
difference is that Estimation-of-Distribution Algorithms try to
``carry over'' the correlations in the population of selected to the
search distribution whereas crossover destroys correlations.  Carrying
over correlations is non-trivial if the search distribution is to be
explorative, i.e., of more entropy: Typically mutation operators add
entropy to the distribution by adding independent noise to each
marginal, but this reduces the mutual information between genes (see
Lemma \ref{lemMut}).

Consider illustration \ref{corr}. In a finite population of 3
individuals, marked by crosses, the values at the 1st and 2nd loci are
correlated (here illustrated by plotting them on the bisecting line).
The crossed population $\CC p$ comprises at most 9 different
individuals; in the special cases $c^{ij}(01)=0$ and
$c^{ij}(01)=\half$ the population is even finite and comprises 3
respectively 9 equally weighted individuals marked by circles.
Mutation adds independent noise, illustrated by the gray shading, to
the alleles of each individual. The two illustrations for the GA
demonstrate that crossover destroys correlations between the alleles
in the initial population instead of carrying it over to the search
distribution: the gray shading is not focused on the bisecting line.
Instead, an EDA would first estimate the distribution of the
individuals in $p$.  Depending on what probabilistic model is used,
this model can capture the correlations between the alleles; in the
illustration the model could be a Gaussian parameterized by the mean
and covariance matrix (just as for the CMA evolution strategy
\cite{hansen-ostermeier:01}) and the estimation of the covariance in
$p$ leads to the highly structured search distribution in which the
entropy of each marginal is increased without destroying the
correlations between them. We capture this difference in the following
definition:
\begin{definition}[Correlated exploration]
  Let $\UU:\, \L^\O \to \L^\O$ be an operator. The following
  conditions need to hold for \emph{almost all} $p\in \O$ which means:
  for all the space $\O$ except for a subspace of measure zero. We
  define
\begin{itemize}
\item $\UU$ is explorative $\iff$ $\D_\UU H > 0$ for almost all
  $p\in\O$,
\item $\UU$ is marginally explorative $\iff$ $\UU$ is explorative
  and $\exists i:~ \D_\UU H^i > 0$ for almost all $p\in\O$,
\item $\UU$ is correlated explorative $\iff$ $\UU$ is explorative and
  $\D_\UU I > 0$, or equivalently $0 < \D_\UU H < \sum_i
  \D_\UU H^i$, for almost all $p\in\O$.
\end{itemize}
\end{definition}

\begin{corollary} From this definition it follows that
\begin{itemize}
\item[a)] If and only if there exist two loci $i$ and $j$ such that
  the marginal crossover mask distribution $c^{ij}(01)$ for these two
  loci is non-vanishing, $c^{ij}(01) = c^{ij}(10) > 0$, then crossover
  $\CC$ is explorative. For every mask distribution $c \in
  \L^{\{0,1\}^N}$, crossover $\CC$ is neither marginally nor
  correlated explorative.
\item[b)] Mutation $\MM$ is marginally but not correlated explorative.
\item[c)] Mutation and Crossover $\MM \circ \CC$ are marginally but not
  correlated explorative.
\item[d)] In the case of a non-trivial genotype-phenotype mapping
  mutation as well as crossover can be phenotypically correlated
  explorative.
\end{itemize}
\end{corollary}
\begin{proof}
  a) That $\CC$ is neither marginally nor correlated explorative
  follows directly from Theorem \ref{theo1}a, which says that for
  every $c \in \L^{\{0,1\}^N}$ and any population $p \in \L^\O$ the
  marginals of the population do not change under crossover, $\D_\CC
  H^i=0$. But under which conditions is $\CC$ explorative?
  
  If, for two loci $i$ and $j$, $c^{ij}(01)$ is non-vanishing, it
  follows that $\CC$ reduces the mutual information between these two
  loci (Theorem \ref{theo2}c). The subspace of populations $p$ that
  do not have any mutual information $I^{ij}$ between these two loci
  is of measure zero. Hence, for almost all $p$, $\D_\CC I^{ij} < 0$
  and, following Theorem \ref{theo1}b this automatically leads to an
  increase of entropy $\D_\CC H^{ij} > 0$ in the compound
  distribution of the two loci and, since $\D_\CC H \ge \D_\CC
  H^{ij}$, also of the total entropy.
  
  The other way around, if, for every two loci $i$ and $j$,
  $c^{ij}(01)$ vanishes it follows that there is no crossover, i.e.,
  on the all-0s and all-1s crossover masks have non-vanishing
  probability. Hence, $\CC = \id$ and is not explorative.
  
  b) In lemma \ref{lemMut} we prove that for every non-uniform
  population $p$ $\D_\MM H>0$, $\D_\MM H^i>0$, and $\D_\MM I<0$.
  
  c) Since both mutation and crossover are not correlative, it
  follows that their composition is also not correlative:
\begin{align*}
\D_\CC I \le 0 ~,~ \D_\MM I \le 0 \quad\To\quad \D_{\MM\CC} I \le 0 \;.
\end{align*}
  
  d) What is different in the case of a non-trivial genotype-phenotype
  mapping? The assumptions we made about the mutation operator
  (component-wise independence) hold only on the genotype space, not
  anymore on the phenotype space: On genotype space mutation kernels
  are product distributions and mutative exploration is marginally
  explorative but not correlated; projected on phenotype space, the
  mutation kernels are in general not anymore product distributions
  and hence phenotypic mutative exploration can correlated.
\end{proof}

\Subsection{crossconclu}
{Conclusion}{}

There are three main points to conclude:
\begin{itemize}
\item First, we point out that crossover does the \emph{inverse} of
  correlated exploration. It destroys correlations in the exploration
  distribution by transforming them into entropy. In an information
  theoretic sense, the exploration distribution after crossover is
  less complex (carrying less mutual information) that before
  crossover. To me it seems just ``countersensible'' to base the
  notion of building blocks on a discussion of crossover. The most
  natural building blocks are individuals carrying the mutual
  information between genes within the exploration distributions.
  Crossover is splitting these building blocks in smaller ones.
  
\item Of course, the crossover exploration distribution can be modeled
  by graphical models since graphical models can model any
  distribution. In that respect, one could certainly design search
  algorithms based on probabilistic models of the search distribution
  (instead of a population) that model crossover GAs---the PBIL is a
  candidate. However, I would challenge to call such an algorithm an
  Estimation-of-Distribution algorithm because its objective is not to
  really estimate the distribution of selected and in particular the
  correlations within this distribution (with the exception of PBIL
  who's objective is to only estimate the marginals which coincides
  with modeling crossover). In general, EDAs go beyond modeling
  crossover since they introduce a quality which is not a quality of
  crossover: correlated exploration.
  
\item Finally, there is a crucial difference between EDAs and
  (crossover) GAs with respect to the self-adaptation of the
  exploration distribution. EDAs always adapt their search
  distribution (including correlations) according to the distribution
  of previously selected solutions. In contrast, the crossover mask
  distribution, that determines where correlations are destroyed or
  preserved, is usually not self-adaptive.
%%  (All though it seems
%%   straight forward, I don't know of any algorithms that include the
%%   mask distribution in the genome, e.g., $\O = \tO \times
%%   \L^{\{0,1\}^N}$, $g = (x,c)$ and then define crossover as
%% \begin{align*}
%% \CC(g|g_0,g_1)
%% = \half \sum_m \x_0(m)\, [\x=\x_0]\, [x=x_0 \otimes_m x_1]\,
%% + \half \sum_m \x_1(m)\, [\x=\x_1]\, [x=x_0 \otimes_m x_1]\;. )
%%   \end{align*}
  However, if considering a phenotypic level (i.e., if we consider an
  indirect encoding) then both, mutational exploration and crossover
  exploration can be correlated and self-adaptive on the phenotypic
  level (see the theory on $\s$-evolution in \cite{toussaint:03},
  realizing that both, the definition of mutation and the definition
  of crossover do not commute with phenotypic equivalence).
\end{itemize}

\hide{
In \cite{toussaint:03} we introduced the notion of the
\emph{$\s$-quality} of an exploration distribution $\s$ defined as
\begin{align*}
- \kld{\s}{F^{(t)}} - H(\s) \comma
F^{(t)} = \frac{\exp f^{(t)}}{\sum_x \exp f^{(t)}(x)} \;,
\end{align*}
where $F^{(t)}$ denotes the exponential fitness distribution (or
`Boltzmann' or `soft-max distribution') of the fitness function
$f^{(t)}$. The $\s$-quality is actually a measure of how well the
exploration distribution matches the distribution of good solutions
taking also the structural properties of these distributions into
account via the Kullback-Leibler divergence. In this language it is
straightforward to discuss the $\s$-quality of a crossover
distribution: If the fitness function is completely decomposable,
i.e.,
\begin{align*}
f^{(t)}(x) = \sum_{i=1}^N f^{(t)}_i(x^i)
\quad\text{or equivalently}\quad
F^{(t)}(x) = \prod_{i=1}^N F^{(t)}_i(x^i) \;,
\end{align*}
then the mutual information $I(F^{(t)})=0$ in the fitness distribution
always vanishes.
}

%\footer
\small
\bibliography{/home/mt/tex/bibs}
\end{document}